\documentclass[singlecol]{epl2}

\newcommand{\be}{\begin{equation}}
\newcommand{\ee}{\end{equation}}
\newcommand{\bea}{\begin{eqnarray}}
\newcommand{\eea}{\end{eqnarray}}

\newcommand{\sech}{\mathrm{sech}}

\usepackage{amsmath,amsfonts,amssymb}
\usepackage{physics}

\title{Neutral-atom qubits in atom-molecular BEC}
\shorttitle{Neutral-atom qubits in BEC} 

\author{Leena Barshilia\inst{1} \and Rajiuddin Sk\inst{1,2} \and Prasanta K. Panigrahi\inst{1,2} \and Avinash Khare\inst{3}}
\shortauthor{L. Barshilia \etal}

\institute{                    
  \inst{1} Department of Physical Sciences, Indian Institute of Science Education and Research Kolkata, West Bengal 741246, India\\
  \inst{2} Center for Quantum Science and Technology, Siksha 'O' Anusandhan University, Bhubaneswar, Odisha 751030, India\\
  \inst{3} Department of Physics, Savitribai Phule Pune University, Pune 411007, India
}

\abstract{Recently, neutral atoms have emerged as a promising platform for quantum computing, offering scalability. In this study, we showcase the realization of atomic qubits in atom-molecular Bose-Einstein condensate, belonging to three distinct classes. In the first case, the condensed molecules form a droplet platform with a flat-top configuration, facilitating effective isolation from both external environments and neighbouring molecules. The second atomic qubits have wavefunctions in the ``pulse" form, exhibiting power law behaviour, whereas the third one has ground and excited state wavefunctions in their respective composite forms, $\sech^2{\beta x}$ and $\sech{\beta x}\tanh{\beta x}$. The localization of the qubits depends on the chemical potential, which is governed by the photo association, providing effective control for qubit manipulation. The relevant parameters, such as energy level separation, healing length, and atom numbers, are found to be influenced by the non-linearity and strength of photo associations governing the behaviour of macroscopic qubits and molecular droplets.}

\begin{document}

\maketitle




\section{Introduction}
In recent years, significant advancements have been achieved in the development of quantum computers, utilizing quantum states and operations to encode and process information. A quantum bit is fragile and very much susceptible to interactions with the environment. Therefore, a good quality qubit with a long coherence time and less sensitive to the environment is the need of the hour. Qubits can be realized through a variety of physical systems. Examples include trapped ions \cite{martinez2016real,figgatt2017complete}, superconducting setups \cite{dicarlo2009demonstration}, quantum dots \cite{loss1998quantum}, and optical processors \cite{zhou2013calculating}. Apart from these physical systems, the array of neutral atoms controlled by beams of light has emerged as a highly potent and scalable technology for manipulating quantum registers containing up to several thousand qubits \cite{saffman2010quantum,saffman2016quantum}. 
Both for superconducting circuits or silicon spin qubits, the challenge lies in manufacturing artificial atoms that are exceedingly difficult to make identical. In contrast, neutral atoms are inherently identical, a crucial factor for enabling efficient quantum computation. In this connection, it is of deep interest to investigate the potential of neutral atoms, which are based on Bose-Einstein condensate (BEC) and readily realizable experimentally \cite{pethick2008bose}. Atomic BEC has been recently investigated for this possibility. The BEC and defect complex have been shown to admit qubit and, in general, qudit states for the defect in the presence of kink solitons for atomic BEC \cite{shaukat2017quantum}.
The dynamics of the two-component BEC are captured by coupled nonlinear Schrödinger equations (NLSE), exhibiting paired solitons \cite{pethick2008bose}. The fact that soliton can realize logical states was proposed by  Laxmanan and Radhakrishnan in optical fibre, governed by an integrable coupled two-component nonlinear Schrödinger equation \cite{radhakrishnan1997inelastic}. As the solitons of this system can have both elastic and non-elastic scattering, these extended objects can be fully controlled, which has been subsequently physically realized.
These dynamics of two-component BEC include a limiting form introduced by Petrov \cite{petrov2001interatomic}, particularly in the context of Bose condensates with Lee-Huang-Yang quantum corrections incorporated via a non-analytic quadratic non-linearity. Bose-Einstein condensates (BECs) can be confined to behave as one-dimensional and two-dimensional systems through a harmonic trap configuration.
The possibility of realizing qubits, with both grey soliton and stable bright soliton, has been considered \cite{steiglitz2000time,javed2022simulating}.
It is worth mentioning that the presence of both dark and bright solitons leads to the well-known Pöschl-Teller potential, which is reflectionless in the appropriate parameter domain and solvable, thereby yielding exact solutions for the qudit system \cite{cooper1995supersymmetry}. In the presence of grey solitons, all-optical controlled-NOT gate implementation has been shown through appropriate reflectionless potential. As is well known, the superconducting qubits underlying most of the present quantum computing platforms are derived from the sine Gordon model, describing the superconducting order parameter \cite{clarke2008superconducting}.

The study of atom-molecular Bose-Einstein condensates (AMBEC) has been a focus of research, especially in the realm of cold chemistry \cite{anderson1995observation,weiner1999experiments}. Cold molecules, with their complex energy structure, hold the potential to drive advancements in quantum engineering and quantum chemistry \cite{bohn2017cold,carr2009cold,quemener2012ultracold}. Due to the difficulty in controlling the cooling of molecules, there has been significant interest in exploring molecular condensation via the atomic route, which begins with an atomic quantum gas and then pairs the atoms into molecules \cite{kohler2006production}. It is of deep interest to investigate the possibility of realizing a quantum computing platform using this atom molecular BEC. These systems have physically attracted attention due to the macroscopic nature of the wavefunctions and robustness against decoherence. Recently, quantum molecular devices have attracted significant attention for quantum information processing due to their compact size, which helps in maintaining coherence over many operations \cite{kosloff2024quantum}.



In this paper, we demonstrate that two orthogonal solutions of atomic and molecular mean field equations can be effectively conceptualized as a ground state and an excited state, thereby forming a qubit system. We find a wide class of exact two-level systems for the AMBEC complex belonging to the atomic level, which may possibly find an application as qubits. These include Pöschl- Teller type solutions, where the molecule can be a quantum droplet and lump-type or bright, dark composite solitons. The spatial spread of power law nature indicates localization in the frequency domain. Furthermore, ``molecular" soliton-type configurations are also found to realize qubits where the molecular condensate is a bright soliton. 
Additionally, we outline the physical operations that can be regarded as quantum gates for quantum computation tasks.

\section{Model}
In recent years, significant advancements have propelled Photoassociation to the forefront of research in the realm of ultracold atomic gases \cite{jones2006ultracold,dastidar2013generation}. This technique involves the formation of molecules from two atoms through the absorption of a photon from an applied optical field during atomic collisions. The development of experimental capabilities has enabled precise quantum control over the intricate internal degrees of freedom within molecules \cite{croft2017universality,krems2008cold,balakrishnan2016perspective}, thereby creating a unique platform for cold and ultracold chemistry.

Originating from early investigations into ultracold atom collisions \cite{weiner1999experiments,julienne1989collisions}, recent efforts have focused on engineering more complex reactions in the quantum regime. Here, phenomena such as quantum threshold laws, quantum statistics, and single partial wave scattering govern collision processes \cite{bell2009ultracold,hutson2007molecular}. These processes are controlled by external electric and magnetic fields.
To describe control mechanism in photoassociation, we focus on diatomic molecular formation, which  represents the most elementary second-order reaction:
\begin{center}
$A + A \Leftrightarrow A_2$.
\label{eq:molecule}
\end{center}
In this scenario, the formation of potential products results in various outcomes as dictated by quantum statistics: $bb\rightarrow b$, $bf\rightarrow f$, and $ff\rightarrow b$, where $b$ represents bosonic and $f$ denotes fermionic counterparts. 

Our focus lies on a chemical system of the former type, where bosonic enhancement exerts the most significant influence on chemical dynamics.
\begin{table}[h!]
	\centering
		\begin{tabular}{c|c|c}
		 Order&Reaction & $\hat{H}_{\text{int}}$ \\ \hline
			0.&$\text{Bath} \Leftrightarrow A$ & $k(\hat{a}_A^\dag+ \text{h.c.})$  \\
			1.&$A \Leftrightarrow B$ & $k(\hat{a}^\dag_{A} \hat{a}_{B} + \text{h.c.})$  \\
			2.&$A + A \Leftrightarrow A_{2}$ & $k(\hat{a}_{A}^\dag\hat{a}_{A}^\dag\hat{a}_{A_2} + \text{h.c.})$  \\
			
		\end{tabular}
		\vspace{3mm}
		\caption{The proposed interaction Hamiltonians \cite{richter2015ultracold} for low-order bosonic reactions.}
			\label{tab:interaction}
\end{table}
In Table \ref{tab:interaction}, various orders of chemical reactions are depicted within a broad context. A zeroth-order reaction represents the exchange of species with a reservoir; on the other hand, a first-order reaction characterizes a linear interaction between two quantum fields. Conversely, atom-molecule interconversion necessitates a Hamiltonian incorporating a second-order reaction.
\begin{equation}
\hat{H} =E_{A}\hat{n}_A +E_{A_2}\hat{n}_{A_2} + k(\hat{a}_A^\dag\hat{a}_A^\dag\hat{a}_{A_2} + h.c.),
		\label{eq:quantum_hamiltonian}
\end{equation}
where $E_{A}$ and $E_{A_2}$ label the corresponding ground-state energies.This Hamiltonian can be extended to encompass the influence of multiple concurrent reactions, particle loss, and dissipation. In our work, we disregard these effects and assume that the reaction rate significantly exceeds the ground state energies, i.e., $k \gg |E_A| + |E_{A_2}|$. It is important to note that within this framework, we describe reversible reactions with a single reaction rate. Furthermore, this model provides insights into the outcome of a chemical reaction without directly revealing the underlying processes of bond breaking and formation.

The system's Hamiltonian can be expressed using field operators for both atoms and the molecular resonant state. When resonance occurs, the quantity of molecules increases significantly, culminating in the creation of a molecular Bose-Einstein condensate (BEC). Our considerations encompass two-body interactions involving atom-atom, molecule-molecule, and atom-molecule collisions, alongside the component accountable for the conversion of pairs of atoms into molecules and vice versa, in terms of natural unit.
\begin{eqnarray}
	\hat{H} &=& \large{\int} {\rm d}^3 r \Big( \hat{\psi}^{\dagger}_a \Big[ -\frac{\hbar^2}{2}\nabla^2 
	+ V_a(\vec r) + \frac{g_a}{2} \hat{\psi}^{\dagger}_a \hat{\psi}_a \Big] \hat{\psi}_a \nonumber \\
	&+& \hat{\psi}^{\dagger}_m \Big[ -\frac{\hbar^2}{4}\nabla^2  
	+ V_m(\vec r) + \epsilon + \frac{g_m}{2} \hat{\psi}^{\dagger}_m \hat{\psi}_m \Big] \hat{\psi}_m \cr \nonumber \\
	&+& g_{am} \hat{\psi}^{\dagger}_a  \hat{\psi}_a  \hat{\psi}^{\dagger}_m  \hat{\psi}_m
	+ \frac{\alpha}{\sqrt{2}} \big[ \hat{\psi}^{\dagger}_m \hat{\psi}_a \hat{\psi}_a 
	+ \hat{\psi}_m \hat{\psi}^{\dagger}_a \hat{\psi}^{\dagger}_a \big] \Big)
	\label{HFR}	
\end{eqnarray}
Here, $g_{a}$, $g_{m}$, and $g_{am}$ quantify the strength of atom-atom, molecule-molecule, and atom-molecule interactions, respectively. The trapping potentials for atoms and molecules are symbolized as $V_a$ and $V_m$, while $\alpha$ denotes the strength of photoassociation (PA). The parameter $\epsilon$ accounts for the energy mismatch during the conversion of atoms to molecules.

In the following, we focus on a cigar-shaped geometry. We maintain the modified parameters from the quasi-one-dimensional geometry \cite{salasnich2002effective, kamchatnov2004dynamics} for ease of notation. The equations of motion for the atomic and molecular mean fields are expressed as:
\begin{eqnarray}
i\frac{\partial\psi_{a}}{\partial t} &=& -\frac{1}{2}\frac{\partial^{2} \psi_{a}}{\partial x^{2}} + (V_a+g_{a}|\psi_{a}|^{2} + g_{am} |\psi_{m}|^{2}) \psi_{a}\nonumber \\ &+& \alpha \sqrt{2}\psi_{m}\psi_{a}^{*}, \label{1} \nonumber\\ \\
i\frac{\partial\psi_{m}}{\partial t} &=& -\frac{1}{4}\frac{\partial^{2} \psi_{m}}{\partial x^{2}}+(V_m+\epsilon +g_{m}|\psi_{m}|^{2}+g_{am}|\psi_{a}|^{2})\psi_{m} \nonumber \\ &+& 
\frac{\alpha}{\sqrt{2}}\psi_{a}^{2} \label{2}. \nonumber\\
\end{eqnarray}
Clearly, when $\alpha$ is non-zero, the total particle count remains conserved,
\begin{equation}
N=\int{(\vert\psi_{a}\vert^2+2\vert\psi_{m}\vert^2}) dx=N_{a}+2N_{m}.
\end{equation}
Here $\psi_{j}$'s are taken as, $\psi_{j}(x,t)=\sqrt{n_{j}(x,t)}e^{i\phi_{j}(x,t)}$ for $j=a,m$., for which the continuity equation can be written as,
\begin{equation}
\frac{\partial}{\partial {t}}(n_{a}+2n_{m})+\frac{\partial}{\partial {x}}(\sum_{a,m}n_{j}\frac{\partial\phi_{j}}{\partial x})=0.
\end{equation}

\section{Quantum droplets}
The coupled AMBEC system, being in general non-integrable, is not amenable to a systematic approach for finding the solution space, unlike the NLSE, where the hierarchy of solutions exists. It is important to note that for the same molecular condensate wave function, multiple atomic wave functions can exist in different parameter domains. Recently extended soliton's solution has been identified \cite{modak2024chemical,modak2022coherent}. Their collisional behaviour has been investigated by P. Das \textit{et al.}\cite{das2016realization}. Here, we demonstrate the realization of atomic qubits with the following three analytic solutions.
The allowed parameter domain and parametric conditions are given in the Supplementary material. These solutions are not only characteristically different but will also have different atom numbers. Physically, the droplets may be thought of as superposed solutions, as will be seen.

We now show that these coupled equations given in Eq. \ref{1} and \ref{2} with the condition $V_{1,2}(x) = 0$ admit three classes of novel solutions, where for the same $\psi_m$, we obtain two solutions in $\psi_a$, one corresponding to the ground state and the other corresponding to the first excited state.

\subsection{ Solution I: Ground state solution in $\psi_a$}
 The exact solution to the coupled 
Eqs. (\ref{1}) and (\ref{2}) has the following form
\be\label{3}
\psi_a = \frac{A\cosh(\beta x)}{B+\cosh^2(\beta x)} e^{-i\mu t}\,,
\ee 
\be\label{4}
\psi_m = \frac{D}{B+\cosh^2(\beta x)} e^{-2i\mu t}\,,
\ee 
provided the desired consistency conditions are satisfied. These are depicted in the Supplementary material.
It is to be noted that $\psi_a$ and $\psi_m$, the spatial wavefunction can be expressed as the following superposition \cite{khare2022superposed}:
$$
\frac{1}{B+\cosh^2(\beta x)} =  \frac{\tanh(\beta x +\Delta)- \tanh(\beta x -\Delta)} {\sinh(2\Delta)},$$
$$
\frac{\cosh(\beta x)}{B+\cosh^2(\beta x)} = \frac{\sech(\beta x+\Delta)+\sech(\beta x -\Delta)}{2\cosh(\Delta)}
$$,
where $B = \sinh^2(\Delta)$.
Physically, this indicates the droplet can be interpreted as a combination of kink and anti-kink, whereas the atomic profile is a superposition of bright solitons.


The solutions can be cast in a form which explicitly shows the range of chemical potential in which the solutions manifest \cite{petrov2001interatomic}.
With $\beta^2 = -2\mu$, we can reexpress the solution given by Eqs. (\ref{3}) and (\ref{4}) as 
\be\label{15}
\psi_a = \frac{2A\cosh(\sqrt{-2\mu} x)}{(2B+1)
[1+\sqrt{1-\frac{\mu}{\mu_0}}\cosh(\sqrt{-8\mu} x)]} e^{-i\mu t}\,,
\ee 
\be\label{16}
\psi_m = \frac{\sqrt{n_m}\frac{\mu}{\mu_0}} 
{[1+\sqrt{1-\frac{\mu}{\mu_0}}\cosh(\sqrt{-8\mu} x)]} e^{-2i\mu t}\,,
\ee 
where
\be\label{17}
\sqrt{n_m} = \frac{D(2B+1)}{2B(B+1)}\,,~~\frac{\mu}{\mu_0} = 
\frac{4B(B+1)}{(2B+1)^2}\,.
\ee

\subsection{Solution II: First excited state solution in $\psi_a$}
The excited atomic state can also be obtained exactly, which has the following form
\be\label{18}
\psi_a = \frac{A\sinh(\beta x)}{B+\cosh^2(\beta x)} e^{-i\mu t}\,,
\ee 
\be\label{19}
\psi_m = \frac{D}{B+\cosh^2(\beta x)} e^{-2i\mu t}\,,
\ee 
provided certain relations are satisfied, which are given in Supplementary material. We note that the atomic excited state profile can also be written in the form 
$$
\frac{\sinh(\beta x)}{B+\cosh^2(\beta x)} = \frac{\sech(\beta x-\Delta) - \sech(\beta x +\Delta)}{2\sinh(\Delta)}$$
 with $B = \sinh^2(\Delta)$, which shows that the superposition of bright solitons with opposite parity leads to the excited state. 



Using $\beta^2 = -2\mu$ , we re-express the solution given by Eqs. (\ref{18}) and (\ref{19}) as
\be\label{30}
\psi_a = \frac{2A\sinh(\sqrt{-2\mu} x)}{(2B+1)
[1+\sqrt{1-\frac{\mu}{\mu_0}}\cosh(\sqrt{-8\mu} x)]} e^{-i\mu t}\,,
\ee 
\be\label{31}
\psi_m = \frac{\sqrt{n_m}\frac{\mu}{\mu_0}} 
{[1+\sqrt{1-\frac{\mu}{\mu_0}}\cosh(\sqrt{-8\mu} x)]} e^{-2i\mu t}\,,
\ee 
where
\be\label{32}
\sqrt{n_m} = \frac{D(2B+1)}{2B(B+1)}\,,~~\frac{\mu}{\mu_0} = 
\frac{4B(B+1)}{(2B+1)^2}\,.
\ee

For both ground state and excited state, the molecular BEC solution, $\psi_m$, has the same form, and $\psi_m$ becomes constant as $\mu\rightarrow\mu_0$. 
\section{Pulse Solution with Power Law Tail}
\subsection{Solution III: Ground state solution in $\psi_a$}

There is another class of solution which shows pulse behaviour. In particular, we can check the exact pulse solution to the coupled Eqs. (\ref{1}) and (\ref{2}) is given by
\be\label{87}
\psi_a = \frac{A}{B+x^2} e^{-i\mu t}\,,
\ee 
\be\label{88}
\psi_m = \frac{D(x^2+y)}{B+x^2} e^{-2i\mu t}\,,
\ee 
where $y$ is any real number with $y \ne B$, provided some consistency conditions given in Supplementary material are satisfied. 

\subsection{Solution IV: First excited state solution in $\psi_a$}

It is easy to check that another exact solution to the coupled Eqs. (\ref{1}) 
and (\ref{2}) is
\be\label{108}
\psi_a = \frac{Ax}{B+x^2} e^{-i\mu t}\,,
\ee 
\be\label{109}
\psi_m = \frac{D(x^2+y)}{B+x^2} e^{-2i\mu t}\,,
\ee 
where $y$ is any real number with $y \ne B$, provided the consistency conditions given in the Supplementary material are satisfied.

\section{Hyperbolic Solution}
\subsection{Solution V: Ground State Solution in $\psi_a$}

We find an exact hyperbolic solution to the coupled equations (\ref{1}) and (\ref{2}), which is
\be\label{4.1}
\psi_a = A\sech^2(\beta x)  e^{-i\mu t}\,,
\ee 
\be\label{4.2}
\psi_m = B [\sech^2(\beta x)+y] e^{-2i\mu t}\,,
\ee 
provided the relations given in the Supplementary material are satisfied.
We have two classes of constraints depending on whether $y=0$ or $y\neq 0$. 

\subsection{Solution VI: First Excited State Solution in $\psi_a$}
We find exact hyperbolic solution  to the coupled Eqs.(\ref{1}) and ((\ref{2})) having the form
\be\label{4.13}
\psi_a = A\sech(\beta x) \tanh(\beta x)  e^{-i\mu t}\,,
\ee 
\be\label{4.14}
\psi_m = B [\sech^2(\beta x)+y] e^{-2i\mu t}\,,
\ee 
provided some relations are satisfied, given in Supplementary material. Depending on whether $y=0$ or $y\neq 0$, two classes of constraints are obtained.
\section{Realization of Quantum Gates}
One-qubit gates are particular unitary transformations represented by 2x2 complex matrices designed to change the state of a single qubit. Examples of such gates include the NOT gate, which flips the state \(|0\rangle\) to \(|1\rangle\) and vice versa, and the Hadamard (H) gate, which transforms a pure state into a superposition of both \(|0\rangle\) and \(|1\rangle\). In the basis \(\{|0\rangle, |1\rangle\}\), these gates are represented as follows:

$$
NOT=\begin{bmatrix}
    0&1\\
    1&0
\end{bmatrix},
H=
\frac{1}{\sqrt{2}}\begin{bmatrix}
1&1\\
    1&-1
\end{bmatrix}
$$

The superposition of macroscopic Bose-Einstein condensate states has been shown both theoretically and experimentally.  In Ref. \cite{ruostekoski1998macroscopic}, authors have shown instead of analyzing spatial interference patterns, one can combine scattered photons from atomic transitions between different BECs using a photon beam splitter.  Simulations show that detecting scattered photons drives the condensates into macroscopic quantum superpositions of phase and number state. This method is theoretically advantageous over atom counting and nondestructive to condensates. In Ref. \cite{kumar2009matter}, it has been shown that matter wave interference patterns can be observed in the intra-trap collision of two bright solitons by selectively tuning the trap frequency and scattering length. The ground and excited atomic states of the AMBEC in the present case differ by parity. We observe that for creating superposition of the atomic molecular BEC states, the angular momentum carrying Lauggere Gaussian beams can be potentially useful \cite{mondal2014angular}. These techniques can be used to make quantum superpositions of BEC, which eventually become the operation of the Hadamard gate. Nevertheless, creating the superposition of these atomic states needs careful consideration.

The other important CNOT gate for carrying out the entangling operation alters the state of a ``target" qubit only when a ``control" qubit is in the state $\ket{1}$. In the context of BEC, Al Khawaja \textit{et al.} introduce a protocol for the quantum controlled-NOT gate, employing a reflectionless potential well in an optical system to manipulate two qubits during soliton scattering \cite{javed2022simulating}. It is worth mentioning that in the context of optical fibre, soliton manipulation for achieving gate operation has been realized. For the integrable  Manakov case, computational universality of the solitons has been established \cite{steiglitz2000time}. Similar methods can be explored in the present case.

\section{Conclusion}
In conclusion, we have explicitly constructed a number of two-level systems with characteristically different macroscopic wavefunctions for the atomic condensate in an atom-molecular BEC. These include hyperbolic $\sinh(\beta x)$ and $\cosh(\beta x)$ type atomic condensate wavefunctions in a molecular droplet background having a flat top characteristic.
These solutions are well-controlled by the photo association and exist in a range of negative chemical potentials. The dark and bright solitonic configurations are also shown to provide a two-level system, similar to the solitons of the Manakov system. Interestingly, in one case, the atomic qubits are produced by the superposition of kink anti-kinks and bright solitons with appropriate parity. In another realization, the atomic qubits are in the molecular form, being a product of bight and kink solitons. The pulse-type solitons are also identified as qubits, which have a Lorentzian density profile, showing power law decay. Expectedly, these spatially extended solutions are well-localized in the momentum domain, providing a characteristically different class of qubit state. We have also discussed the procedure for creating energy level superposition to realize the Hadamard gate, as well as the possible realization of the CNOT gate. The detailed investigation of gate operations, as well as scalability \cite{hokmabadi2019supersymmetric}, are under investigation.

\end{document}


\maketitle
For convenience, we are restating the equations of motion for the atomic and molecular mean fields with potential $V=0$:
\begin{eqnarray}
i\frac{\partial\psi_{a}}{\partial t} &=& -\frac{1}{2}\frac{\partial^{2} \psi_{a}}{\partial x^{2}} + (g_{a}|\psi_{a}|^{2} + g_{am} |\psi_{m}|^{2}) \psi_{a}\nonumber \\ &+& \alpha \sqrt{2}\psi_{m}\psi_{a}^{*}, \label{1} \nonumber\\ \\
i\frac{\partial\psi_{m}}{\partial t} &=& -\frac{1}{4}\frac{\partial^{2} \psi_{m}}{\partial x^{2}}+(\epsilon +g_{m}|\psi_{m}|^{2}+g_{am}|\psi_{a}|^{2})\psi_{m} \nonumber \\ &+& 
\frac{\alpha}{\sqrt{2}}\psi_{a}^{2} \label{2}. \nonumber\\
\end{eqnarray}

\textbf{(A) Solution I}

We use the ansatz in Eq. (\ref{2}), which yields the following relations satisfied by different coefficients
\be\label{5}
(2\mu -\epsilon)D = -D\beta^2 +\frac{\alpha A^2}{\sqrt{2}}\,,
\ee
\be\label{6}
g_{am} A^2 = (2\mu-\epsilon)B -\frac{(3+4B)}{2}\beta^2\,,
\ee
\be\label{7}
g_m D^2 = (2\mu -\epsilon)B^2 + \frac{B}{2}\beta^2
\ee

On the other hand, on using the ansatz in Eq. (\ref{1}) yields the following three relations
\be\label{8}
\mu = -\frac{\beta^2}{2}\,,~~g_a A^2 + \sqrt{2} \alpha D = -(1+4B)\beta^2\,,
\ee
\be\label{9}
g_{am} D^2 - B g_a A^2 = 4B(B+1)\beta^2\,.
\ee
Without any loss of generality, we put $D^2 = \eta A^2$ where $\eta > 0$ in these relations, and we obtain the following relations, which have a simpler form 
\be\label{10}
\beta^2 = -2\mu\,,~~\epsilon = -[\frac{(1+2B)\eta +2B}{2\eta B}] \beta^2\,,
\ee
\be\label{11}
\alpha = \frac{2B+(1+2B)\eta}{\sqrt{2} BD}\beta^2\,,
\ee
\be\label{12}
g_{am} A^2 = -\frac{(1+2B)\eta -B}{\eta}\beta^2\,,
\ee
\be\label{13}
g_{a} A^2 = -\frac{(1+2B)\eta+2+3B+4B^2}{B}\beta^2\,,
\ee
\be\label{14}
g_{m} A^2 = \frac{B(B+\eta)\beta^2}{\eta^2}\,.
\ee
{\bf (B) Solution II}
On using the ansatz in Eq. (\ref{2})
yields three relations
\be\label{20}
(2\mu -\epsilon)D = -D\beta^2 +\frac{\alpha A^2}{\sqrt{2}}\,,
\ee
\be\label{21}
g_{am} A^2 = (2\mu-\epsilon)(B+1) -\frac{(1+4B)}{2}\beta^2\,,
\ee
\be\label{22}
g_m D^2 = (2\mu -\epsilon)(B+1)^2 - \frac{(B+1)\beta^2}{2}\,.
\ee

On the other hand, on using the ansatz in Eq. (\ref{1}) yields the three 
relations
\be\label{23}
\mu = -\frac{\beta^2}{2}\,,~~g_a A^2 + \sqrt{2} \alpha D = -(3+4B)\beta^2\,,
\ee
\be\label{24}
g_{am} D^2 - (1+B) g_a A^2 = 4B(B+1)\beta^2\,.
\ee

Without any loss of generality, let $D^2 = \eta A^2$ where $\eta > 0$. In
terms of $\eta$ the relations (\ref{20}) to (\ref{24}) take slightly simpler
form
\be\label{25}
\beta^2 = -2\mu\,,~~\epsilon = [\frac{2(1+B)-(2B+1)\eta}{2(B+1)\eta}]\beta^2\,,
\ee
\be\label{26}
\alpha = -\frac{[2(1+B)-(1+2B)\eta]}{\sqrt{2} D(B+1)}\beta^2\,,
\ee
\be\label{27}
g_{am} A^2 = -\frac{(1+2B)\eta +(B+1)}{\eta}\beta^2\,,
\ee
\be\label{28}
g_{a} A^2 = -\frac{(1+2B)(\eta-1)}{(B+1)}\beta^2\,,
\ee
\be\label{29}
g_{m} A^2 = -\frac{(1+B)(\eta+ B+1)}{\eta^2}\beta^2\,.
\ee

{\bf (C) Solution  III}
On using the ansatz in Eq. (\ref{1}) yields three relations
\be\label{89}
\mu = g_{am} D^2 + \sqrt{2} \alpha D\,,
\ee
\be\label{90}
2(B-y) g_{am} D^2  +\sqrt{2} \alpha D (B-y) =-3\,,
\ee
\be\label{91}
(B^2-y^2) g_{am} D^2 + \sqrt{2} \alpha D B(B-y) = B +g_a A^2\,.
\ee

On the other hand, on using the ansatz in Eq. (\ref{2}) yields the four 
relations
\be\label{92}
2\mu - \epsilon = g_{m} D^2 = 0\,,
\ee
\be\label{93} 
(3/2)(B-y)+ g_{am} A^2 +\frac{\alpha A^2}{\sqrt{2}D} = 0\,,
\ee
\be\label{94}
-(B/2)(B-y)+ y g_{am} A^2 +\frac{\alpha A^2 B}{\sqrt{2}D}=0\,.
\ee

On solving Eqs. (\ref{93}) and (\ref{94}) we obtain
\be\label{95}
g_{am} A^2 = -2B\,,~~\frac{\alpha A^2}{\sqrt{2}D} = (B+3y)/2\,.
\ee

On using $D^2 = \eta A^2$ and comparing Eqs. (\ref{90}) and (\ref{95})
we then obtain
\be\label{96}
\eta = \frac{1}{(B-y)^2}\,,~~\mu = \frac{\epsilon}{2} = \frac{3y-B}
{(B-y)^2}\,,~~g_a A^2 = -2B\,.
\ee

{\bf (D) Solution IV}
On using the ansatz in Eq. (\ref{1}) yields three relations
\be\label{110}
\mu =  g_{am} D^2 + \sqrt{2} \alpha D\,,
\ee
\be\label{111}
(B^2-y^2) g_{am} D^2  +\sqrt{2} \alpha D B (B-y) = 3B\,,
\ee
\be\label{112}
(B-y)^2 g_{am} D^2 = B g_a A^2 -4B\,.
\ee

On the other hand, on using the ansatz in Eq. (\ref{2}) yields the four 
relations
\be\label{113}
2\mu - \epsilon = g_{m} D^2 = -\frac{B}{2y(B+y)}\,,~~
\frac{\sqrt{2}\alpha y A^2}{D} = -(B+3y)\,,
\ee
\be\label{114} 
2y (B+y) g_{am} A^2  = (-B^2+6By+3y^2)\,.
\ee

On using $D^2 = \eta A^2$ in Eqs. (\ref{112}) and (\ref{114}) we then
find that
\be\label{115}
\eta = -\frac{2By}{(B-y)^2(B+y)}\,,~~g_a A^2 = \frac{(5B^2+y^2+2By)}
{(B+y)^2}\,.
\ee
It is now straight forward to compute $\mu$ and $\epsilon$ for this 
solution.

{\bf (E) Solution V}
On using the ansatz in Eq. (\ref{1}) we find that either $y = 0$ or $y \ne 0$. We will discuss both cases separately, one by one.

{\bf Case I: $y \ne 0$}

In that case from Eq. (\ref{1}) we obtain the following three relations
\be\label{4.3}
\mu = -2\beta^2 + g_{am} B^2 y^2  +\sqrt{2} \alpha B y\,,
\ee
\be\label{4.4}
g_{am} B^2  = -g_a A^2\,,
\ee
\be\label{4.5}
 2y g_{am} B^2  +\sqrt{2} \alpha B = -3 \beta^2\,,
\ee

On the other hand, on using the ansatz in Eq. (\ref{2}) yields the  
relations
\be\label{4.6}
g_{am} A^2  = -g_{m} B^2\,,
\ee
\be\label{4.7}
2\mu - \epsilon = g_m B^2 y^2 = \frac{\beta^2}{2}\,,
\ee
\be\label{4.8}
\frac{\alpha A^2}{\sqrt{2}B} = -\frac{(3y+2)\beta^2}{2y}\,,
\ee

On assuming $B^2 = \eta A^2$ where obviously $\eta > 0$, on using
Eqs. (\ref{4.6}) and (\ref{4.8}) in Eq. (\ref{4.5}) we find that 
\be\label{4.9}
\eta = \frac{y}{y+1}\,.
\ee
Since $\eta > 0$, this implies that either $y > 0$ or $y < -1$.

{\bf Case II: $y = 0$}

In this case it is easily checked that while from Eq. (\ref{1}) we have
relation (\ref{4.4}) and further
\be\label{4.10}
\mu = -2\beta^2\,,~~\sqrt{2} \alpha B = -3\beta^2\,,
\ee

On the other hand from Eq. (\ref{2}) we obtain the relation (\ref{4.6}) 
and further
\be\label{4.11}
\frac{\alpha A^2}{\sqrt{2}B} = -\frac{3\beta^2}{2}\,,~~2\mu -\epsilon 
= -\beta^2\,.
\ee
On comparing Eqs. (\ref{4.10}) and (\ref{4.11}) it follows that 
\be\label{4.12}
B = \pm A\,,~~g_{am} = -g_{a} = -g_{m}\,,~~\alpha A 
= \pm \frac{3\beta^2}{\sqrt{2}}\,.
\ee

{\bf (F) Solution VI}
 On using the ansatz in Eq. (\ref{1})
we find that either $y = 0$ or $y \ne 0$. We will discuss both cases 
seperately one by one.

{\bf Case I: $y \ne 0$}

In that case from Eq. (\ref{1}) we obtain the following three relations
\be\label{4.15}
\mu = -\frac{\beta^2}{2} + g_{am} B^2 y^2  +\sqrt{2} \alpha B y\,,
\ee
\be\label{4.16}
g_{am} B^2  = g_a A^2\,,
\ee
\be\label{4.17}
 (1+2y) g_{am} B^2  +\sqrt{2} \alpha B = -3 \beta^2\,,
\ee

On the other hand, on using the ansatz in Eq. (\ref{2}) yields the  
relations
\be\label{4.18}
g_{am} A^2  = g_{m} B^2\,,
\ee
\be\label{4.19}
2\mu - \epsilon = g_m B^2 y^2\,,
\ee
\be\label{4.20}
\beta^2 = \frac{\alpha A^2}{\sqrt{2}B}+g_{am} A^2 y +2g_{m} B^2 y^2\,,
\ee
\be\label{4.21}
-\frac{3\beta^2}{2} = -\frac{\alpha A^2}{\sqrt{2}B} +(1-y) g_{am} A^2 
+3y g_{m} B^2\,.
\ee

On assuming $B^2 = \eta A^2$ where obviously $\eta > 0$, 
we find that 
\be\label{4.22}
(2y+1)g_{am} A^2 = -[2(1+y)+\frac{(3+2y)}{\eta}] \beta^2\,,
\ee
\be\label{4.23}
\frac{\alpha A^2}{\sqrt{2} B} = [(1+y)+\frac{y}{\eta}]\beta^2\,.
\ee

{\bf Case II: $y = 0$}

In this case it is easily checked that from Eq. (\ref{1}) we have
relation (\ref{4.16}) and further
\be\label{4.24}
\mu = -\frac{\beta^2}{2}\,,~~\sqrt{2} \alpha B + g_{am} B^2 = -3\beta^2\,,
\ee

On the other hand from Eq. (\ref{2}) we obtain the relation (\ref{4.18}) 
and further
\be\label{4.25}
\frac{\alpha A^2}{\sqrt{2}B}-g_{am} A^2 = \frac{3\beta^2}{2}\,.
\ee
\be\label{4.26}
2\mu -\epsilon = -\beta^2 +\frac{\alpha A^2}{\sqrt{2}B}\,,
\ee

On using $B^2 = \eta A^2$ where $\eta > 0$, from Eqs. (\ref{4.24}) and 
(\ref{4.25}) we obtain
\be\label{4.27}
g_{am} A^2 = -\frac{(1+\eta)\beta^2}{\eta}\,,~~\frac{\alpha A^2}{\sqrt{2} B}
= \frac{(\eta-2)\beta^2}{2\eta}\,,
\ee
\be\label{4.27}
g^2_{am} = g_{a}g_{m}\\
\ee